\documentclass[ twocolumn,08 pt, amsmath,amssymb, aps,fleqn]{revtex4-1}
\usepackage[hidelinks]{hyperref}
\usepackage[english]{babel}
\usepackage{mathtools}
\usepackage{tabularx}
\usepackage{multirow}
\usepackage{comment}
\usepackage{graphicx}
\usepackage{color}
\usepackage{soul}
\begin{document}
\preprint{APS/123-QED}
\title{The Quantum Steeplechase}
\author{Joy Prakash Das}  \author{Chandramouli Chowdhury} \author{Girish S. Setlur}\email{gsetlur@iitg.ernet.in}
\affiliation{Department of Physics \\ Indian Institute of Technology  Guwahati \\ Guwahati, Assam 781039, India}
\begin{abstract}
Quantum Steeplechase is the study of a Luttinger liquid (LL) in one dimension in the presence of a finite number of barriers and wells clustered around an origin. The powerful non-chiral bosonization technique (NCBT) is introduced to write down closed formulas for the two-point functions in the sense of the random phase approximation (RPA). Unlike g-ology based methods that are tied to the translationally invariant, free particle basis, the NCBT explicitly makes use of the translationally non-invariant single particle wavefunctions. The present method that provides the most singular part of the asymptotically exact Green function in a closed form, is in contrast to competing methods that require a combination of renormalization group and/or numerical methods in addition to the bosonization techniques.
\end{abstract}

\maketitle
\section{Introduction}
In quantum many-body physics, the stated goal is to write down all the ``N-point Green functions" of a system of many mutually interacting particles in the thermodynamic limit. The N-point particle (hole) Green function  is the quantum overlap between two states of a system where each state has N particles added (removed) at various locations and times. An analytical study (as against a numerical one) of mutually interacting quantum particles is beset with formidable technical difficulties and various approximation techniques are used to mitigate these problems. The obvious method that springs to mind is to expand in powers of the interaction potential between the quantum particles.  In one dimension, each term in this perturbation series carried out in momentum space, diverges logarithmically at low momenta (known as infra-red divergences). Hence a ``non-perturbative" method is called for. For translationally invariant systems, this method, which goes under the name `g-ology' is well established (see e.g. Giamarchi \cite{giamarchi2004quantum}). The g-ology method is tied to the translationally invariant free particle basis and the `restricted Hilbert space of states', in which the Fermi-Bose correspondence is justified, makes even the study of free fermions in the presence of barriers/wells (or weak links) quite formidable. By contrast, the present approach which amounts to constructing the `restricted Hilbert space of states' not for free fermions but for free fermions plus these barriers/wells or weak links, is able to study the problem of Luttinger liquids in the presence of these imperfections more easily and is able to provide analytical expressions for the most singular part of the Green functions and so on that interpolate between the weak barrier and weak link cases.

\begin{figure}[h!]
\begin{center}
\includegraphics[scale=0.25]{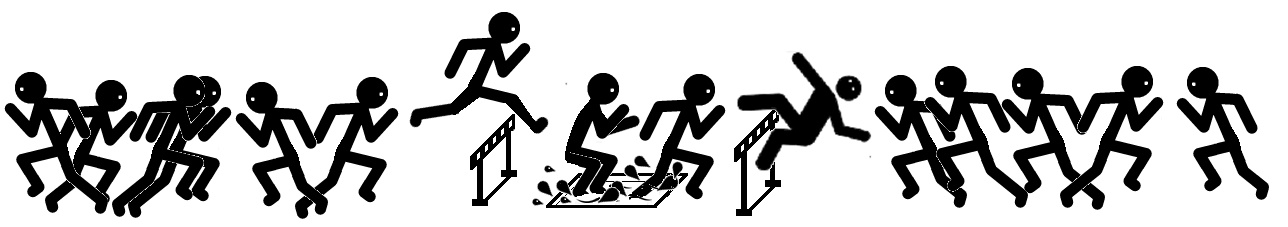}
\end{center}
\caption{ The Quantum Steeplechase: Athletes (representing electrons) crossing/bouncing off hurdles (potential barriers) and water-jumps (potential wells) while moving in both directions with the fastest athlete possessing the Fermi momentum and rubbing shoulders with each other (representing forward scattering short-range interactions)}
\label{steeplechase}
\label{quantum_steeplechase}
\end{figure}
The study of the effect of impurities in Luttinger liquids constitutes an important area of theoretical condensed matter physics, especially in the last few decades. The detailed study of transport in Luttinger liquid (LL) in the presence of a weak link was started by Kane and Fisher \cite{kane1992transport} followed by the study of a LL near a double barrier \cite{kane1992resonant}. Since then a number of papers have appeared that have generalised these ideas using a variety of approaches which include fermionic renormalization \cite{meden2002single}, path integral approaches \cite{fernandez2001friedel}, functional renormalization \cite{meden2003scaling,andergassen2004functional,enss2005impurity,gezzi2007functional,jakobs2007nonequilibrium}, flow equations for Hamiltonians \cite{stauber2003tomonaga}, functional integral formalism \cite{grishin2004functional}, Monte Carlo methods \cite{hamamoto2008numerical} and so on. Different physical phenomena are also studied in Luttinger liquids with impurities: Friedel oscillations \cite{Egger1995friedel1}, conductance \cite{fendley1995exact, fendley1995exact2}, Kondo effect \cite{furusaki1994kondo, schiller1995exact}, etc.
Experimental realizations of 1D systems gives a motivational boost to study quantum physics in one dimension. In this regard, Luttinger liquid behavior in carbon nano-tubes \cite{bockrath1999Luttinger,egger2000Luttinger}, experimental evidences of resonant tunneling in a Luttinger liquid \cite{auslaender2000experimental} are worth mentioning.



 But what is missing in the existing literature are explicit  expressions of the correlation functions of a Luttinger liquid with localized potentials of arbitrary strengths in terms of elementary functions of positions and times. The best available are limiting cases for a weak barrier \cite{giamarchi2004quantum} and an infinite barrier \cite{mattsson1997properties} which can be obtained using conventional bosonization schemes. The goal of this work is the introduction of a simple, appealing  but powerful analytical tool: the Non chiral bosonization technique, using which we are able to provide the most singular part of the asymptotically exact Green functions of a strongly inhomogeneous Luttinger liquid for arbitrary interaction strengths of the impurity and with the short-range  forward scattering between the fermions. 

The paper is organized as follows. Section II describes the problem and summarizes the solution procedure. Section III contains the calculation of the Green function for free fermions, which forms the basis for the bosonization technique. The subsequent section describes the Non chiral bosonization technique followed by the results in Section V. Various limiting cases are verified and compared favorably with existing literature. The last section summarizes how the technique can be used to study other types of systems and also the various physical phenomena which can be explained using the Green functions obtained using NCBT.


\section{Problem description}
Consider a Luttinger liquid in one dimension with forward scattering short-range mutual interactions \cite{giamarchi2004quantum} in the presence of a scalar potential V(x) that is localized near an origin. The full  generic-Hamiltonian of the system(s) under study (before taking the RPA limit) is (are),
\small
\begin{equation}
\begin{aligned}
H =& \int^{\infty}_{-\infty} dx \mbox{    } \psi^{\dagger}(x) \left( - \frac{1}{2m} \partial_x^2 + V(x) \right) \psi(x)\\
  & \hspace{1cm} + \frac{1}{2} \int^{ \infty}_{-\infty} dx \int^{\infty}_{-\infty} dx^{'} \mbox{  }v(x-x^{'}) \mbox{   }
 \rho(x) \rho(x^{'})
\label{Hamiltonian}
\end{aligned}
\end{equation}
\normalsize
where $ v(x-x^{'}) = \frac{1}{L} \sum\limits_{q}  v_q \exp{[ -i q(x-x^{'})] } $ (where $ v_q = 0 $ if $ |q| > \Lambda $ for some fixed bandwidth $ \Lambda \ll k_F $ and $ v_q = v_0 $ is a constant, otherwise) is the forward scattering mutual interaction. Also, $ V(x) $ is the external potential which represents the cluster of impurities around a fixed point. The following potentials have been considered.\\
\begin{equation}
\label{potentials}
V(x)=
\begin{cases}
&V_0 \delta(x)\mbox{ }\hspace{2.7 cm}(\text{single delta})\\
&V_0( \delta(x+a)+\delta(x-a))\mbox{ } \mbox{ }(\text{double delta})\\
&V_1 \delta(x+a)+V_2 \delta(x-a) \mbox{ }  \mbox{ }(\substack{\text{asymmetric} \\\text{double delta}})\\
& V_0 \delta(x) +  V_1(\delta(x\pm a))\mbox{ }\mbox{ }\mbox{ }\mbox{ }(\text{triple delta})\\
&V \theta(x+a)\theta(a-x)\hspace{1 cm}(\text{finite barrier})\\
&\hspace{-0.2 cm}-V \theta(x+a)\theta(a-x)\hspace{0.9 cm}(\text{finite well})\\
\end{cases}\\
\end{equation}

Here $\theta(x)$ is the Heaviside step function. The density is given by $ \rho(x,t) = \psi^{\dagger}(x,t) \psi(x,t) - \rho_0 $ (no point splitting is required before taking RPA limit). The central goal of this paper is to write down the Green functions of these systems at zero and at finite temperature in the presence of the potentials described in eq. (\ref{potentials}) . For an analytical solution to be feasible when mutual interactions are included, it is necessary to confine the study to the so-called RPA limit which means, among other things, working in the limit where the Fermi momentum and the mass of the fermion diverge in such a way that their ratio is finite (i.e. $ k_F, m \rightarrow \infty $ but $ k_F/m = v_F < \infty  $: units that make $ \hbar = 1 $, so that $ k_F $ is both the Fermi momentum as well as a wavenumber, are used) \cite{stone1994bosonization}. This amounts to linearizing the energy momentum dispersion near the Fermi surface ($E=E_F+p v_F$ instead of $E=p^2/(2m)$). Furthermore, if `w' is the width of the cluster, it is then imperative to define how $ w $ scales in the RPA limit. The assertion made is that in the RPA limit $ k_F w   < \infty $ as $ k_F \rightarrow \infty $. Similarly the heights and depths of the various barriers are assumed to be in fixed ratios with the Fermi energy $ E_F = \frac{1}{2} m v_F^2 $ even as $ m \rightarrow \infty $ with $ v_F < \infty $.

The goal of calculating the correlation functions is achieved through the following steps which are elaborated in the subsequent sections.
{\bf a)} First the single particle two-point functions are calculated in the RPA limit in presence of the cluster of impurities. 
{\bf b)} From the two-point functions, the slow part of the density density correlation function (DDCF) is calculated. 
{\bf c)} The two-point functions in (a) are expressed in terms of the densities calculated in (b), which is now called the bosonized version of the Green function. 
{\bf d)}  The DDCFs in (b) is modified to include interactions.
{\bf e) } In the bosonized version of the Green function obtained in (c), all the densities are replaced by their interacting versions obtained in (d) to obtain the Green function in presence of interactions.\\

\section{Green's functions of free fermions}
Denote the full two-point Green function (also known as single particle Green function) of the system before taking the RPA limit (i.e. with parabolic energy-momentum relation) as \small $
 <T\mbox{  }\psi(x,\sigma,t)\psi^{\dagger}(x',\sigma',t')> $ \normalsize where the time ordering decides whether it is particle or hole Green function that is being studied and $ \sigma $ is the spin projection of the individual fermions. In terms of this, the asymptotic or RPA Green function is defined by ``smearing out" the positions and times over the scale of the Fermi wavelength and Fermi times as follows,
\small
\begin{equation}
\begin{aligned}
\label{TWOPOINT}
\hspace*{-0.3 cm}
\langle &T\mbox{  }\psi_{\nu}(x,\sigma,t)\psi_{\nu'}^{\dagger}(x',\sigma',t')\rangle  = \lim_{m \rightarrow \infty }\\
& \ll \langle T\psi(y,\sigma,\tau)\psi^{\dagger}(y',\sigma',\tau')\rangle e^{-ik_F(\nu y - \nu' y')}e^{ i E_F (\tau-\tau^{'}) } \gg\\
\end{aligned}
\end{equation}
\normalsize
where,
\begin{equation}
\begin{aligned}
\hspace{1cm}&\ll f(t) \gg \mbox{  } = \mbox{  }  \frac{1}{2T_F} \int^{ t+T_F}_{t-T_F} d\tau \mbox{   }f(\tau)\hspace{1 cm}\\
&\ll g(x) \gg \mbox{  } = \mbox{  } \frac{1}{2\lambda_F} \int^{ x+\lambda_F}_{x-\lambda_F} dy \mbox{   }g(y) \
\end{aligned}
\end{equation}
\normalsize
with \footnotesize$ \lambda_F = 2\pi/k_F $ \normalsize  and \footnotesize $ T_F =  2\pi/E_F $,   $ k_F = m v_F $ \normalsize and \footnotesize $ E_F =  (1/2) m v_F^2 $
with \footnotesize $ v_F < \infty $  \normalsize being held fixed. Also, here $ \nu,\nu^{'} = \pm 1 $ correspond to the right and left Fermi points.

We start with the non-interacting Hamiltonian (after dropping the last term of eq. (\ref{Hamiltonian})) and calculate the two orthonormal set of wavefunctions, one with the propagation starting from the left and the other with that starting from the right. The wavefunctions are subjected to the RPA limit discussed in the previous section. Using the spectral decomposition method \cite{de1993exact, andrade2014exact} the two-point Green function is obtained in position and energy coordinates, which undergoes a Fourier transform to yield the space-time two-point functions which has the following form (at zero temperature),
\normalsize
\begin{equation}
\begin{aligned}
\hspace*{-0.2 cm}
\label{INPUT1}
&\langle T\mbox{  }\psi_{\nu}(x,\sigma,t)\psi_{\nu'}^{\dagger}(x',\sigma',t')\rangle_0\\
&\hspace{1 cm}= \sum_{\gamma,\gamma'=\pm 1}\frac{ \theta(\gamma x)\theta(\gamma' x')\mbox{ }g_{\gamma,\gamma'} (\nu,\nu')}{(\nu x-\nu' x')-v_F(t-t')}\delta_{\sigma,\sigma'}
\end{aligned}
\end{equation}
where $ \theta(x) $ is Heaviside's step function and $ g_{\gamma,\gamma'} (\nu,\nu') $ are given in table \ref{gvalue}.

\noindent\begin{table}[h!]
\caption{  Values of  $ g_{\gamma,\gamma'} (\nu,\nu') $ for the general case.  The explicit expressions for T's and R's for the subcases are given in \hyperref[AppendixA]{Appendix A} . }
{\begin{tabular}{c c c c c c}
\hline
&$g_{(\gamma,\gamma')}(\nu,\nu')$  &$\substack{\gamma=1\\\gamma'=1}$&$\substack{\gamma=-1\\\gamma'=-1}$&$\substack{\gamma=1\\\gamma'=-1}$&$\substack{\gamma=-1\\\gamma'=1}$\\	[5pt]
\hline
&$(\nu,\nu')=(1,1)$ &\hspace{0.1 cm}$\frac{i}{2\pi}$\hspace{0.1 cm} &\hspace{0.1 cm}$\frac{i}{2\pi}$\hspace{0.1 cm}
 &\hspace{0.1 cm}$\frac{i}{2\pi}T$\hspace{0.1 cm} &\hspace{0.1 cm}$\frac{i}{2\pi}T^*$\hspace{0.1 cm}\hspace{0.1 cm}\\	[4pt]
&$(\nu,\nu')=(-1,-1)$ & $\frac{i}{2\pi}$&$\frac{i}{2\pi}$ & $\frac{i}{2\pi}T^{*}$&$\frac{i}{2\pi}T$\\	[4pt]
&$(\nu,\nu')=(1,-1)$ & $\frac{i}{2\pi}R$& $\frac{i}{2\pi}R^{*}$&0 &0\\	[4pt]
&$(\nu,\nu')=(-1,1)$ & $\frac{i}{2\pi}R^{*}$& $\frac{i}{2\pi}R$& 0&0\\	[4pt]
\hline
\end{tabular}
}
\label{gvalue}
\end{table}


\normalsize
The values of the transmission($T $) and reflection($ R $) amplitudes are calculated for all the sub-cases of eq. (\ref{potentials}) and they are given in \hyperref[AppendixA]{Appendix A}.  Note that for potentials which lack inversion symmetry about any chosen point, (e.g. asymmetric double deltas) the presence of nontrivial phases in $ T $ and $ R$ contribute to the expected lack of inversion symmetry in the Green's functions.

Note that in eq. (\ref{INPUT1}), the term $ [(\nu x-\nu' x')-v_F(t-t')] $ appears in the denominator. In general, in a Luttinger liquid with mutual interactions, this term appears with a non-trivial system dependent exponent viz. as $ [ (\nu x-\nu' x')-v_F(t-t') ]^g $. Listing these $ g $'s and other similar exponents  is one of the main goals of this paper since $ g = 1 $ is only when mutual interaction between fermions are absent. It is easy to generalize these results to finite temperature since for this a simple replacement, viz., $ \frac{1}{X} \rightarrow \frac{\pi}{ \beta v_F} \mbox{  } csch[ \frac{\pi X}{\beta v_F} ] $ is sufficient where e.g. $ X  \equiv [(\nu x-\nu' x')-v_F(t-t')] $ and $ \beta $ is inverse temperature.\\
\subsection{Density density correlation function}
The other main goal of this paper to write down the density-density correlation function (DDCF) of the system which is a special case of a 4-point function.  In the RPA sense, the density $ \rho(x,t) $ may be ``harmonically analysed" as follows.
\begin{equation}\label{INPUT2}
\rho(x,t) = \rho_s(x,t) + e^{ 2 i k_F x } \mbox{   }\rho_f(x,t) +  e^{ - 2 i k_F x } \mbox{   }\rho^{*}_f(x,t)
\end{equation}
The slowly varying part of the density $ \rho_s $ (the average density is subtracted out, so this is really the deviation) has an auto-correlation function which when mutual interactions are absent, may be written down using Wick's theorem as follows,

\footnotesize
\begin{equation}
\begin{aligned}\label{INPUT3}
\langle T \mbox{    }\rho_s(x,t)\rho_s(x^{'},t^{'})\rangle_0=-\sum_{ \substack{\gamma,\gamma^{'} \\= \pm 1}}\hspace{0.2 cm}\sum_{ \substack{ \nu,\nu^{'}\\=\pm 1}} \frac{ |g_{\gamma,\gamma^{'}}(\nu,\nu^{'})|^2 \mbox{  } \theta(\gamma x) \theta(\gamma^{'} x^{'})
 }{ [(\nu x - \nu^{'} x^{'}) - v_F (t-t^{'}) ]^2 }
\end{aligned}
\end{equation}
\normalsize
where $ g_{\gamma,\gamma'} (\nu,\nu') $ are given in table \ref{gvalue}.
These three relations viz.  eq. (\ref{INPUT1}), eq. (\ref{INPUT2}) and   eq. (\ref{INPUT3}) shall be used in the subsequent sections as input to the NCBT scheme in order to enable an explicit evaluation of the Green functions.

\section{Bosonized version of the N-point Green's functions}
Just as the density may be harmonically analysed, the field may  also be harmonically analysed so that  $ \psi(x) =  e^{i k_F x }  \mbox{  }\psi_R(x)  + e^{-i k_F x }  \mbox{  }\psi_L(x)  $. 
 Bosonization  involves inverting the defining formulas for current and number densities viz. $ j(x) = Im[\psi^{\dagger}(x)\partial_x\psi(x)] $ and  $ \rho(x)  = \psi^{\dagger}(x)\psi(x) $ and rewriting $  \psi(x) $ in terms of $ j $ and $ \rho $. Then the continuity equation  $ \partial_t \rho + \partial_x j = 0 $ is invoked to write $ \psi(x) $ purely as a (non-local) function of $ \rho $ and $ \partial_t \rho $. It follows therefore, that the  the N-point function is some combination of the correlations of the density field with itself. Bosonization may be thought of as the ``inverse of Wick's theorem". While Wick's theorem - which is valid only for systems with no mutual interactions - seeks to express higher order correlations in terms of lower order ones, bosonization seeks to express the single particle Green function in terms of the higher order density-density correlations.
The inversion of the defining relation between current and densities in the standard bosonization scheme that goes by the name g-ology (see the book by Giamarchi \cite{giamarchi2004quantum}) yields the following relation between $ \psi_{\nu}(x,\sigma,t) $ (where $ \nu = R(+1) \mbox{  }or\mbox{   } L(-1) )$ and the slowly varying part of the density (this is a mnemonic for generating the N-point functions),
\begin{equation}
\begin{aligned}
\psi_{\nu}(x,\sigma,t) \sim \exp{\big[ i \theta_{\nu}(x,\sigma,t) \big]}
\label{PSINU}
\end{aligned}
\end{equation}
with the local phase given by the formula,
\small
\begin{equation}
\begin{aligned}
\theta_{\nu}(x,\sigma,t) =& \pi \int^{x}_{sgn(x)\infty} dy \bigg( \nu  \mbox{  } \rho_s(y,\sigma,t)\\
&\hspace{1 cm} -  \int^{y}_{sgn(y)\infty} dy^{'} \mbox{ }\partial_{v_F t }  \mbox{ }\rho_s(y^{'},\sigma,t) \bigg)
\end{aligned}
\end{equation}\normalsize
The above prescription in eq. (\ref{PSINU}) is valid for nearly translationally invariant systems (i.e. with possible external potentials with Fourier components small compared to the Fermi momentum) and for systems with a half line (no tunneling across the barrier). 
In the present approach, a modification of the correspondence of eq. (\ref{PSINU}) is introduced wherein the correlation functions of a system of free fermions plus barriers and wells with arbitrary heights (depths) can be as easily computed in the bosonized language as it is in the original Fermi language.
\noindent
\begin{equation}
\begin{aligned}
\psi_{\nu_i}&(x_i,\sigma_i,t_i) \rightarrow  \sum_{\gamma_i = \pm 1}\sum_{ \lambda_i \in \{0,1\} }C_{\lambda_{i}  ,\nu_i,\gamma_i}(\sigma_i)\mbox{ }
\theta(\gamma_i x_i) \mbox{ }\\
&\hspace{0.7 cm}e^{ i \theta_{\nu_i }(x_i,\sigma_i,t_i) + 2 \pi i \nu_i \lambda_{i}  \int^{x_i}_{sgn(x_i)\infty} \rho_s(-y_i,\sigma_i,t_i) \mbox{  }dy_i }
\label{PSII}
\end{aligned}
\end{equation}
\normalsize
This  `non-standard harmonic analysis' is an alternative to the usual one invoked while using g-ology methods which is valid for translationally invariant systems and half lines whereas the harmonic analysis in eq. (\ref{PSII}) is valid for systems considered in this paper. Current algebra, point splitting constraints, etc. continue to be obeyed by this new Fermi-Bose correspondence.

An analogy with the anharmonic oscillator problem in undergraduate quantum mechanics may be useful. One could either study this problem in the the translationally invariant plane wave basis or more conveniently in a basis closer to the real ground state of the system viz. the states of the simple harmonic oscillator.
While technically there is nothing wrong with using the plane wave basis, this really makes the problem quite complicated. Using eq. (\ref{PSINU}) to study the problem of fermions in presence of barriers and wells is somewhat like using the plane wave basis to study the anharmonic oscillator. It is much better to use eq. (\ref{PSII}) which is analogous to using the harmonic oscillator basis to study the anharmonic oscillator.

The quantities  $ C_{\lambda_{i}  ,\nu_i,\gamma_i}(\sigma_i) $ are c-numbers and involve cutoffs and such, which, as in the traditional approach, are not obtainable using these techniques. The only quantities that have absolute meaning are the anomalous exponents i.e., numbers $ g $ when the term involved appears as $ [(\nu x - \nu^{'}x^{'}) - v_F (t-t^{'})]^g $.  The operators that appear in the exponent in eq. (\ref{PSII}) are the ones that are really crucial in this approach since they  provide the right anomalous exponents.  The crucial new ingredient in the modified formula in eq. (\ref{PSII}) is the term involving $  \rho_s(-y_i,\sigma_i,t_i)  $ that ensures that the effects of backscattering from the external potentials are automatically and naturally taken into account so that the mandated trivial exponents are obtained when eq. (\ref{PSII}) is used to compute the N-point functions in the sense of the RPA. The addition of these new terms does not spoil fermion commutation rules since there is a prefactor of $ 2 \pi i \nu_i $ next to it which ensure that fermion commutation relations of the fields are respected. These new terms also do not spoil the point-splitting constraints for the Fermi bilinears, which is an opaque way of saying that when eq. (\ref{PSII}) is used to infer the currents and densities - as the latter two are, after all, bilinears of the Fermi fields - the resulting expressions are in accordance with expectations. 

In order to extract the anomalous exponents of the system with mutual interactions, two things remain. One is to generalize eq. (\ref{INPUT3}) to include mutual interactions.
The other is to derive a prescription for choosing the values of the crucial parameters $ \lambda_i = 0,1 $ which indicates when the traditional form of the field needs modification.  It simply involves making sure that the prescription (which is unique) leads to N-point functions of the system (without mutual interactions) identical to what is given by Wick's theorem.
This is done subsequently below. In addition to these $ \lambda_i's $, auto-correlation functions of the slowly varying parts of the density when mutual interactions are present are needed.

Again in the spirit of the RPA, the density density correlation functions given in eq. (\ref{INPUT3}) are modified to include mutual interactions and the following formula may be obtained
  ($ \rho_h(x,t) =  \rho_s(x,\uparrow,t) + \rho_s(x,\downarrow,t)  $ is the ``holon" density and $ \rho_{n}(x,t) =  \rho_s(x,\uparrow,t) - \rho_{s}(x,\downarrow,t)  $ is the ``spinon" density and $ a = h $ for holon and $ a = n $ for spinon) 
\footnotesize
\begin{equation}
\begin{aligned}
\langle T\mbox{   } \rho_a(x_1,t_1)\rho_a(x_2,t_2)\rangle  = &\frac{v_F  }{ 2\pi^2 v_a } \mbox{   } \sum_{  \nu = \pm 1 }\bigg (   \frac{-1}{ ( x_1-x_2 + \nu v_a(t_1-t_2) )^2 }\\
&	-  \frac{\frac{v_F }{v_a}  \mbox{    } \text{sgn}(x_1) \text{sgn}(x_2)\mbox{   }Z_a}{  ( | x_1|+|x_2 | + \nu v_a(t_1-t_2) )^2 }
\bigg)
\label{RHOSRHOS}
\end{aligned}
\end{equation}\normalsize
where $ a = n$ (spinon) or  h (holon) and,
\begin{equation}
 Z_a = \frac{ |R|^2 }{    \bigg( 1 - \delta_{a,h} \frac{(v_h-v_F)}{ v_h }
 |R|^2   \bigg) }
 \label{Za}
\end{equation}
\normalsize
Here the spinon velocity is just the Fermi velocity since it is the total density that couples to the short-range potential:
 $ v_n = v_F $, but the holon velocity is modified by interactions, \scriptsize $ v_h = \sqrt{v_F^2+2v_F v_0/\pi} $ \normalsize where the interaction between fermions is the two-body short-range forward scattering potential which just means the potential between two particles at $ x $ and $ x^{'} $ is \footnotesize  $ V(x-x^{'}) = \frac{1}{L}\sum_{|q| < \Lambda }v_0 \mbox{ } \exp{[ -i q (x-x^{'})] } $, \normalsize   where $ \Lambda $ is held fixed as the RPA limit is taken. Finally, $ \langle T \mbox{ }\rho_n(x_1,t_1)\rho_h(x_2,t_2)\rangle \equiv 0 $.  It can be shown that an expansion of eq. (\ref{RHOSRHOS}) in powers of $ v_0 $ matches with the corresponding series obtained by standard perturbation theory so long as one retains only the most singular terms.

\section{ Full two-point Green's function}
The two-point (single-particle) Green's function may be written down  using the correspondence in eq. (\ref{PSII}).  Only the anomalous exponents which refer to the constants $ g $ that appear in terms of the form $ [ \nu_1 x_1 - \nu_2 x_2 - v_F(t_1-t_2)]^g $  that emerge from this calculation are of interest here. These $ g $'s are uniquely pinned down once a prescription for deciding which of the $ \lambda_i's $ are zero or one and under what circumstances is given. This prescription follows unambiguously by requiring that an evaluation (of the 2M-point function)  in the Gaussian (and RPA) sense leads to trivial exponents when mutual interactions between fermions are absent. Of lesser importance are the coefficients $ C's $ which  depend on the details of the potentials and cutoffs and other such non-universal features, as is also the case in the conventional approach. The prescription for obtaining the $ \lambda_i's $ are simple. Consider a general 2M-point function. Imagine  mentally pairing up one annihilation operator with one creation operator and create M such pairs. This is simply a mental activity since this pairing (Wick's theorem) is not valid when mutual interactions are present.  Consider one such pair and let the two $ \lambda $'s of this pair be  $  (\lambda_m,\lambda_k) $ where $ k > m $. The constraints are as follows:

\vspace{-0.4 cm}
\footnotesize
\begin{equation}
\lambda_m = \begin{cases} \lambda_k &\mbox{if } (\nu_m,\nu_k) = (\gamma_m,\gamma_k) \mbox{   }\mbox{or} \mbox{   }(\nu_m,\nu_k) = (-\gamma_m,-\gamma_k) \\
1-\lambda_k &\mbox{if } (\nu_m,\nu_k) = (-\gamma_m,\gamma_k) \mbox{   }\mbox{or}\mbox{   }(\nu_m,\nu_k) = (\gamma_m,-\gamma_k)\end{cases}
\label{PRES}
\end{equation}
\normalsize
\vspace{-0.4 cm}

 This (unique) prescription guarantees the right trivial exponents in the right places when mutual interactions are turned off. The full Green's function in presence of interactions are as follows ({\bf Notation:} \begin{small}$X_i \equiv (x_i,\sigma_i,t_i)$, \end{small} also, in order to remove ambiguities associated with cutoff dependent quantities in translationally non-invariant systems, the notion of weak equality denoted by \begin{small} $ A[X_1,X_2] \sim B[X_1,X_2] $ \end{small} is introduced which really means  \begin{small} $ \partial_{t_1} Log[ A[X_1,X_2] ]  = \partial_{t_1} Log[ B[X_1,X_2] ] $\end{small} assuming of course, A and B do not vanish identically. Furthermore the finite temperature versions of the formulas below are obtained by replacing $ Log[Z] $ by $ Log[ \frac{\beta v_F }{\pi}Sinh[ \frac{\pi Z}{ \beta v_F} ] ] $ where $ Z \sim  (\nu x_1 - \nu^{'} x_2 ) - v_a (t_1-t_2)  $ and singular cutoffs ubiquitous in this subject are suppressed in this notation for brevity - they have to be understood to be present. The notion of weak equality is unable to pin down possible prefactors in the Green functions that may even  be spatially inhomogeneous (but time independent) in addition to being singular. The inhomogeneous prefactors are nothing but
  terms such as $ e^{ \frac{1}{2}\langle A^2 \rangle } $ and $  e^{ \frac{1}{2} \langle B^2 \rangle } $ that come about when
  when evaluating $ \langle e^A e^B \rangle = e^{ \frac{1}{2} \langle A^2 \rangle }  e^{ \frac{1}{2} \langle B^2 \rangle } e^{ \langle AB \rangle } $ when $ \langle AB \rangle \mbox{  } \propto \mbox{   }Log[(\nu x_1 - \nu^{'} x_2 ) - v_a (t_1-t_2)] $. It must be stressed that these inhomogenous prefactors are important for extracting the exponents associated with tunneling conductance and the local dynamical density of states. Here $\tau_{12} =  t_1 - t_2$):\\

\begin{bf} Case I : $x_1$ and $x_2$ on the same side of the origin\end{bf} \\ \scriptsize

\begin{equation*}
\begin{aligned}
\Big\langle T\mbox{  }\psi&_{R}(X_1)\psi_{R}^{\dagger}(X_2)\Big\rangle \sim 
\frac{(4x_1x_2)^{\gamma_1}}{(x_1-x_2 -v_h \tau_{12})^{P} (-x_1+x_2 -v_h \tau_{12})^{Q}} \\
\times&\frac{1}{ (x_1+x_2 -v_h \tau_{12})^{X} (-x_1-x_2 -v_h \tau_{12})^{X} (x_1-x_2 -v_F \tau_{12})^{0.5}}\\
\Big\langle T\mbox{  }\psi&_{L}(X_1)\psi_{L}^{\dagger}(X_2)\Big\rangle \sim 
\frac{(4x_1x_2)^{\gamma_1}}{(x_1-x_2 -v_h \tau_{12})^{Q} (-x_1+x_2 -v_h \tau_{12})^{P}} \\
\times&\frac{1}{ (x_1+x_2 -v_h \tau_{12})^{X} (-x_1-x_2 -v_h \tau_{12})^{X}(-x_1+x_2 -v_F \tau_{12})^{0.5}}\\
\Big\langle T\mbox{  }\psi&_{R}(X_1)\psi_{L}^{\dagger}(X_2)\Big\rangle \sim 
\frac{(2x_1)^{\gamma_1}(2x_2)^{1+\gamma_2}+(2x_1)^{1+\gamma_2}(2x_2)^{\gamma_1}}{2(x_1-x_2 -v_h \tau_{12})^{S} (-x_1+x_2 -v_h \tau_{12})^{S}} \\
\times&\frac{1}{ (x_1+x_2 -v_h \tau_{12})^{Y} (-x_1-x_2 -v_h \tau_{12})^{Z}(x_1+x_2 -v_F \tau_{12})^{0.5}}\\
\end{aligned}
\end{equation*}

\begin{equation}
\begin{aligned}
\Big\langle T\mbox{  }\psi&_{L}(X_1)\psi_{R}^{\dagger}(X_2)\Big\rangle \sim 
\frac{(2x_1)^{\gamma_1}(2x_2)^{1+\gamma_2}+(2x_1)^{1+\gamma_2}(2x_2)^{\gamma_1}}{2(x_1-x_2 -v_h \tau_{12})^{S} (-x_1+x_2 -v_h \tau_{12})^{S}} \\
\times&\frac{1}{ (x_1+x_2 -v_h \tau_{12})^{Z} (-x_1-x_2 -v_h \tau_{12})^{Y}(-x_1-x_2 -v_F \tau_{12})^{0.5}}\\
\label{SS}
\end{aligned}
\end{equation}

\small
\begin{bf}Case II : $x_1$ and $x_2$ on opposite sides of the origin\end{bf} \\ \scriptsize

\begin{equation}
\begin{aligned}
\Big\langle T\mbox{  }\psi&_{R}(X_1)\psi_{R}^{\dagger}(X_2)\Big\rangle \sim 
\frac{(2x_1)^{1+\gamma_2}(2x_2)^{\gamma_1} }{2(x_1-x_2 -v_h \tau_{12})^{A} (-x_1+x_2 -v_h \tau_{12})^{B}} \\
\times&\frac{(x_1+x_2)^{-1}(x_1+x_2 + v_F \tau_{12})^{0.5}}{ (x_1+x_2 -v_h \tau_{12})^{C} (-x_1-x_2 -v_h \tau_{12})^{D} (x_1-x_2 -v_F \tau_{12})^{0.5}}\\
&\hspace{2cm}+\frac{(2x_1)^{\gamma_1} (2x_2)^{1+\gamma_2}}{2(x_1-x_2 -v_h \tau_{12})^{A} (-x_1+x_2 -v_h \tau_{12})^{B}} \\
\times&\frac{(x_1+x_2)^{-1}(x_1+x_2 - v_F \tau_{12})^{0.5}}{ (x_1+x_2 -v_h \tau_{12})^{D} (-x_1-x_2 -v_h \tau_{12})^{C} (x_1-x_2 -v_F \tau_{12})^{0.5}}\\
\Big\langle T\mbox{  }\psi&_{L}(X_1)\psi_{L}^{\dagger}(X_2)\Big\rangle \sim 
\frac{(2x_1)^{1+\gamma_2}(2x_2)^{\gamma_1} }{2(x_1-x_2 -v_h \tau_{12})^{B} (-x_1+x_2 -v_h \tau_{12})^{A}} \\
\times&\frac{(x_1+x_2)^{-1}(x_1+x_2 - v_F \tau_{12})^{0.5}}{ (x_1+x_2 -v_h \tau_{12})^{D} (-x_1-x_2 -v_h \tau_{12})^{C} (-x_1+x_2 -v_F \tau_{12})^{0.5}}\\
&\hspace{2cm}+\frac{(2x_1)^{\gamma_1} (2x_2)^{1+\gamma_2}}{2(x_1-x_2 -v_h \tau_{12})^{B} (-x_1+x_2 -v_h \tau_{12})^{A}} \\
\times&\frac{(x_1+x_2)^{-1}(x_1+x_2 + v_F \tau_{12})^{0.5}}{ (x_1+x_2 -v_h \tau_{12})^{C} (-x_1-x_2 -v_h \tau_{12})^{D} (-x_1+x_2 -v_F \tau_{12})^{0.5}}\\
\Big\langle T\mbox{  }\psi&_{R}(X_1)\psi_{L}^{\dagger}(X_2)\Big\rangle \sim \mbox{ }0\\
\Big\langle T\mbox{  }\psi&_{L}(X_1)\psi_{R}^{\dagger}(X_2)\Big\rangle \sim  \mbox{ }0\\
\label{OS}
\end{aligned}
\end{equation}

\normalsize
The analytical expressions of the anomalous exponents in eq. (\ref{SS}) and eq. (\ref{OS}) are listed in section \hyperref[LuttingerExpo]{Anomalous exponents}. The highlight of this work are the formulas described in Case II above. It is easy to see that even after setting $|R| = 0$, these Green functions do not correspond to the translationally invariant Luttinger liquid. This implies that even a small reflection coefficient changes the properties of the system drastically when the two points are on opposite sides of the origin. The two-point functions in eq. (\ref{SS}) and eq. (\ref{OS}) obey the Schwinger-Dyson equation and they agree to those obtained by standard fermionic perturbation theory \cite{das2018non}.
\subsection{Anomalous exponents}
\label{LuttingerExpo}

The explicit expressions of the anomalous exponents that appeared in eq. (\ref{SS}) and eq. (\ref{OS}) are listed below.

\footnotesize

\begin{equation}
Q=\frac{(v_h-v_F)^2}{8 v_h v_F} \mbox{ };\mbox{ }  X=\frac{|R|^2(v_h-v_F)(v_h+v_F)}{8  v_h (v_h-|R|^2 (v_h-v_F))}  \mbox{ };\mbox{ }C=\frac{v_h-v_F}{4v_h}
\label{luttingerexponents}\end{equation}
\normalsize
The other exponents can be expressed in terms of the above exponents.
\footnotesize
\begin{equation*}
\begin{aligned}
&P= \frac{1}{2}+Q  \mbox{ };\hspace{0.8 cm}    S=\frac{Q}{C}( \frac{1}{2}-C)   \mbox{ };\hspace{0.85 cm}      Y=\frac{1}{2}+X-C  ;           \\
& Z=X-C\mbox{ };\hspace{0.8 cm}      A=\frac{1}{2}+Q-X \mbox{ };\hspace{0.8 cm}   B=Q-X  \mbox{ };\hspace{1 cm}   \\
&D=-\frac{1}{2}+C   \mbox{ };\hspace{.6 cm}      \gamma_1=X                \mbox{ };\hspace{1.65 cm}    \gamma_2=-1+X+2C;\\
\end{aligned}
\end{equation*}
\normalsize
Many who work in this field are puzzled by two features of our approach and results. The first is that the Luttinger exponents depend on the reflection coefficient ($|R|$) of the cluster of barriers and wells. The other is the fact we continue to use the bare reflection and transmission coefficients in the final formulas whereas in other approaches these are scale dependent.
The first puzzle is easier to clear up. Physically speaking, the Green functions for a homogeneous LL ($|R|=0$) will only contain translational terms $(\pm(x_1-x_2)- v(t_1-t_2))$ and not reflectional terms $(\pm(x_1+x_2)- v(t_1-t_2))$. But for a half line ($|R|=1$) both types will be present. It is the $|R|$ dependence of the Luttinger exponents which will tune this accordingly.

With regard to the reflection coefficients being the bare ones in our approach we have to point out that the scale dependence in the conventional approaches comes about either because the starting point is far removed from the actual situation or because curvature effects, etc. in the free fermion dispersion are not neglected. The former case is when one tries to study a LL in presence of impurity by treating the impurity as a perturbation or study it by treating it as a weak coupling between two half lines. In both cases the various parameters are likely to be scale dependent due to the poor choice of the starting situation in comparison with the actual situation. To give an analogy, if one tries to study the harmonic oscillator Green function using perturbative RG where the spring constant is a perturbation (analogous to treating impurity as a perturbation), it is naturally going to be scale dependent as it is a relevant perturbation. Conversely if one tries to model this as a sequence of particle in a box with weak coupling between boxes (analogous to a weak link between two half-lines), here too the couplings are going to flow.
The present work on the other hand, treats the impurity exactly and strictly neglects the curvature of the free fermion energy dispersion. We also restrict ourselves to forward scattering interaction between fermions. Even with all these qualifications and caveats our results are only able to provide the most singular part of the asymptotic Green functions in a closed form in terms of elementary  functions of positions and times. This is the most important physics that is of interest and it is gratifying that it may be obtained exactly.
\normalsize

\subsection{Limiting case checks}
{\bf No interaction. }
The obvious limiting check is to switch off the inter-particle interactions between particles ( $v_0 = 0$ ) and then compare with the respective single particle Green functions obtained using Fermi algebra. In such a case, the holon velocity is equal to the Fermi velocity ($v_h \to v_F$)  and eq. (\ref{SS}) and eq. (\ref{OS}) will be identical to eq. (\ref{INPUT1})

{\bf No impurity. }
In absence of any impurity, there is no reflection ($|R|=0$) and no concept of opposite sides as its a homogeneous case. There will be no reflectional terms like $\langle\psi_{R}\psi_{L}^{\dagger}\rangle$ and $\langle\psi_{L}\psi_{R}^{\dagger}\rangle$ as obvious from table (\ref{gvalue}). The only non zero terms are the translational terms $\langle\psi_{R}\psi_{R}^{\dagger}\rangle$ and $\langle\psi_{L}\psi_{L}^{\dagger}\rangle$ whose exponents takes the following form.\small
\begin{equation*}
P=\frac{(v_h+v_F)^2}{8v_hv_F};\mbox{ }Q=\frac{(v_h-v_F)^2}{8v_hv_F};\mbox{ }X=\gamma_1=0;
\end{equation*}
\normalsize
Using the above, one obtains the precise Green functions of the standard homogeneous Luttinger liquid as given in books by Giamarchi \cite{giamarchi2004quantum}.

{\bf No tunneling. }
In this case $R=-1$ (half-line) and hence there is no need to consider the two points to be on the opposite sides. The Green functions take the form that of an infinite barrier for the points on the same side and vanishes when one of the points is at the location of the impurity. The Green functions of a half line are calculated by Mattsson et al. \cite{mattsson1997properties} for small values of interaction parameter which are in conformity with those obtained using NCBT subjected to the same conditions.

{\bf Far from impurity. }
It can be observed that when both the points are on the same side of the impurity but far away from it, then the translational terms $\langle\psi_{R}\psi_{R}^{\dagger}\rangle$ and $\langle\psi_{L}\psi_{L}^{\dagger}\rangle$ are immune to the presence of impurity and takes the form of the homogeneous case. But the reflectional terms $\langle\psi_{R}\psi_{L}^{\dagger}\rangle$ and $\langle\psi_{L}\psi_{R}^{\dagger}\rangle$ will certainly not be immune to the presence of the impurity since in these cases the region where the impurity is present needs to be traversed.

\subsection{Spinless case}
The Green functions in eq. (\ref{SS}) and (\ref{OS}) can be easily converted to the corresponding spinless case. All one needs to do is to double all the holon exponents, viz., P to 2P, Q to 2Q, $\gamma_1$ to 2$\gamma_1$, $(1+\gamma_2)$ to 2$(1+\gamma_2)$, etc. and let all the spinon exponents vanish (0.5, 0, -0.5, etc. are set to zero). Thus there will be only one modified velocity given by $v_h=\sqrt{v_F^2+v_F v_0/\pi}$ indicating no spin-charge separation.

\section{Applications}
The formalism described can also be successfully used to obtain the correlation functions of a fermionic one step ladder system \cite{das2017one} and that of a Luttinger liquid with slowly moving impurities \cite{das2018ponderous}. The method can be used to calculate the four point functions relevant to the study of Friedel oscillations in the class of systems described in this work \cite{das2018friedel}. The two-point correlation functions obtained can be used to study some important physical phenomena like resonant tunneling, conductance, dynamical density of states \cite{das2018friedel}, etc. It is worth mentioning that standard phenomena like `cutting the chain' and `healing the chain' described in the seminal paper by Kane and Fisher \cite{kane1992transport} can be elucidated using the conductance study from these Green functions \cite{das2018transport}.

\section{Conclusions}
 In this work, the formalism of the non-chiral bosonization technique (NCBT) has been laid down which is an alternative to the conventional g-ology based methods, especially for systems that are inhomogeneous, as the former is capable of providing the most singular parts of the asymptotically exact Green functions of such systems as closed analytical expressions without using RG methods. 
These formulas interpolate between the weak barrier and weak link extreme cases that are studied in the literature. Unlike the competing methods that can only study these extreme limits reliably, the present approach is able to connect the two regimes using analytical means.  The results thus obtained match fully with various obvious as well as non-trivial limiting cases of the corresponding Green functions found in the literature. The obtained Green functions, which can be validated using perturbation theory, Schwinder-Dyson equation, etc., can be used to study important physical phenomena like conductance, Friedel oscillations and so on.
\section*{Acknowledgement}
 Our deepest gratitude to Prof. Duncan Haldane for sharing his valuable views and opinions. We would also like to thank former research scholar of IIT Guwahati, Dr. V. Meera for her prior collaboration. We also express our sincere gratitude to Dr. Amarendra Sarma, Dr. Diptiman Sen,  Dr. Navinder Singh, Dr. G. Baskaran and Dr. Amit Kumar Agarwal for their useful inputs and suggestions.  \mbox{ }

 \section*{Funding}
   A part of this work was done with financial support from Department of Science and Technology, Govt. of India DST/SERC: SR/S2/CMP/46 2009.\\

\section*{APPENDIX A:  Transmission and Reflection Amplitudes}
\label{AppendixA}
\setcounter{equation}{0}
\renewcommand{\theequation}{A.\arabic{equation}}

\noindent The amplitude reflection and transmission amplitudes for various situations are listed below. These are in turn used to write explicit expressions for the Green function of the translationally non-invariant system when mutual interactions between fermions are absent. These latter results are then used as inputs in the NCBT formalism. The transmission and reflection amplitudes of the six cases shown in eq. (\ref{potentials}) are shown below:

\vspace{0.2in}

\scriptsize
\begin{bf}(a) Single delta-function \end{bf}
\begin{equation}
\begin{aligned}
T=&\frac{1}{\left(1+V_0 \frac{i}{v_F}\right)}\mbox{ };\mbox{ }
R=-\frac{iV_0}{v_F\left(1+V_0 \frac{i}{v_F}\right)} \\
\end{aligned}
\end{equation}

\begin{bf}(b) Symmetric double delta-function \end{bf}
\begin{equation}
\begin{aligned}
T=&\frac{1}{\left(1+V_0 \frac{i}{v_F}\right)^2-\left(\frac{i V_0}{v_F}e^{i \xi_0}\right)^2}\\
R=&-\frac{2i\frac{V_0^2}{v_F^2} \sin{[\xi_0]} +\frac{2i V_0}{v_F}\cos{[\xi_0]}}{\left(1+V_0 \frac{i}{v_F}\right)^2-\left(\frac{i V_0}{v_F}e^{i \xi_0}\right)^2} \\
\end{aligned}
\end{equation}

\begin{bf}(c) Asymmetric double delta-function \end{bf}
\begin{equation}
\begin{aligned}
T=&\frac{1}{\left(1+i\frac{V_1+V_2}{v_F}+\frac{i^2 V_1V_2}{v_F^2}\right)+\frac{V_1V_2}{v_F^2}e^{2 i \xi_0}}\\
R=&-\frac{2 i \frac{V_1 V_2}{v_F^2} \sin{[\xi_0]}+\frac{2i}{v_F}(\frac{V_1 e^{i \xi_0}+V_2e^{-i\xi_0}}{2})}{\left(1+i\frac{V_1+V_2}{v_F}+\frac{i^2 V_1V_2}{v_F^2}\right)+\frac{V_1V_2}{v_F^2}e^{2 i \xi_0}}
\end{aligned}
\end{equation}

\begin{bf}(d) Symmetric triple delta-function \end{bf}
\begin{equation}
\begin{aligned}
T=&\frac{1}{\splitfrac{\left(1-i\frac{V_0V_1^2}{v_F^3}-2\frac{V_0V_1}{v_F^2}-\frac{V_1^2}{v_F^2}+i\frac{V_0}{v_F}+2i\frac{V_1}{v_F}\right)}{+\frac{e^{i\xi_0}}{v_F^2}\left(2i\frac{V_0V_1^2}{v_F}-ie^{i\xi_0}\frac{V_0V_1^2}{v_F}+2V_0V_1+e^{i\xi_0}V_1^2\right)}}\\\\
R=&-\frac{\splitfrac{2i\frac{V_0V_1^2}{v_F^3}-2i \frac{V_0V_1^2}{v_F^3}\cos{[\xi_0]}+2i\frac{V_0V_1}{v_F^2}\sin{[\xi_0]}+}{2i\frac{V_1^2}{v_F^2}\sin{[\xi_0]}+i\frac{V_0}{v_F}+2i\frac{V_1}{v_F}\cos{[\xi_0]}}}{\splitfrac{\left(1-i\frac{V_0V_1^2}{v_F^3}-2\frac{V_0V_1}{v_F^2}-\frac{V_1^2}{v_F^2}+i\frac{V_0}{v_F}+2i\frac{V_1}{v_F}\right)}{+\frac{e^{i\xi_0}}{v_F^2}\left(2i\frac{V_0V_1^2}{v_F}-ie^{i\xi_0}\frac{V_0V_1^2}{v_F}+2V_0V_1+e^{i\xi_0}V_1^2\right)}} \\
\end{aligned}
\end{equation}
\small
\noindent Note: $\lambda=\frac{V}{E_F} $ is fixed while  taking the RPA limit and $ V $ is the well depth or barrier height and $ E_F = \frac{1}{2} mv_F^2 $ is the Fermi energy. $\xi_0=2 k_F a$ where $k_F$ is the Fermi momentum and the barrier(or well) goes from `-a' to `a'.\\ \mbox{  }\\ \mbox{  }\\
\scriptsize
\begin{bf}(e) Finite barrier tunneling \end{bf}
\begin{equation}
\begin{aligned}
T=& \frac{4ie^{-i\xi_0}\sqrt{\lambda-1}}{4 i \sqrt{\lambda -1} \cosh{[\xi_0\sqrt{(\lambda-1)}]}+2(2-\lambda)\sinh{[\xi_0\sqrt{(\lambda-1)}]}}\\
R=&\frac{e^{-i\xi_0}2\lambda\sinh{[\xi_0\sqrt{(\lambda-1)}]}}{4 i \sqrt{\lambda -1} \cosh{[\xi_0\sqrt{(\lambda-1)}]}+2(2-\lambda)\sinh{[\xi_0\sqrt{(\lambda-1)}]}} \\
\end{aligned}
\end{equation}

\begin{bf}(f) Finite well \end{bf}
\begin{equation}
\begin{aligned}
T=&\frac{4e^{-i\xi_0}\sqrt{\lambda+1}}{4  \sqrt{\lambda +1} \cos{[\xi_0\sqrt{(\lambda+1)}]}-2i(2+\lambda)\sin{[\xi_0\sqrt{(\lambda+1)}]}} \\
R=& \frac{e^{-i\xi_0}2i\lambda\sin{[\xi_0\sqrt{(\lambda+1)}]}}{4  \sqrt{\lambda +1} \cos{[\xi_0\sqrt{(\lambda+1)}]}-2i(2+\lambda)\sin{[\xi_0\sqrt{(\lambda+1)}]}}
\end{aligned}
\end{equation}
\normalsize
It can be shown that on taking proper limiting conditions, one can obtain one case from another, for example, from finite barrier to singe delta, from asymmetric double delta to symmetric double delta and so on. \\


\bibliographystyle{apsrev4-1}
\bibliography{ref}
\normalsize

\end{document}